\documentclass[aps,twocolumn,showpacs,amsmath,amssymb]{revtex4}
\usepackage{dcolumn}
\usepackage{bm}
\usepackage{color}
\usepackage{latexsym}
\usepackage{amssymb}
\usepackage{graphicx}
\usepackage{ulem}

\begin{document}
\title{Two-dimensional electron gas as a sensitive noise detector}

\author{M. V. Cheremisin}

\affiliation{A.F.Ioffe Physical-Technical Institute, 194021
St.Petersburg, Russia}

\begin{abstract}
The dc voltage observed at low-temperatures in a 2D electron sample $without$ external excitation is accounted by the Schottky contact rectification of the noise generated in the measuring circuit. The rectified voltage is shown to depend on the asymmetry of the contact pair. The dependence of the rectified voltage on the noise amplitude first follows the trivial quadratic law, then exhibits a nearly linear behavior, and, finally, levels-off.
\end{abstract}


\maketitle

\section{\label{sec:Introduction}Introduction}
In the present paper we examine the experimental setup well known for routine low-T transport measurements.
Let a 2D electron gas( 2DEG ) sample be placed( Fig.1) in a sample chamber kept at liquid-helium temperature. The current leads are attached to the sample, and then connected to external measuring terminal kept at room temperature. The measuring circuit is connected to the terminal. Without external excitation, the dc voltmeter connected to arbitrary 2DEG sample contacts demonstrates [1]-[5] a puzzling nonzero voltage(NV) on the order of $\sim \mu$V. The value and the sign of the dc potential depends on a actual contact pair.

In the presence of a magnetic field, the measured dc potential demonstrates strong( $\sim$mV ) oscillations named "zero" oscillations ZO, which exhibit a $1/B$-periodicity similar to the well known Shubnikov-de Haas(SdH) oscillations. The ZO period allows one to extract the carrier density, whereas the temperature dependence of the ZO amplitude is similar to that for SdH oscillations and gives the correct value of the carrier effective mass. In contrast to SdH oscillations, ZO are skew-symmetric. The amplitude and the phase shift of the zero oscillations depend on a chosen contact pair. Moreover, for a certain contact pair, the ZO shape is strongly affected when other sample leads are connected to(disconnected from) the measuring circuit[6]. We emphasize that the effects observed are, in general, universal and are observed in various 2DEG systems and for arbitrary sample configuration.

The basic idea put forward in Refs.[3,5] in order to explain the NV and ZO consists in the possible rectification of the input noise by the 2D-3D Schottky diodes formed at the sample contacts. As demonstrated in Refs.[3,5], a careful screening of the circuit diminishes the amplitude of the rectified voltage. Then, shunting of the sample contacts by a capacitance suppresses [5] the dc potential as well. In order to quantitatively examine the influence of the noise, both the voltmeter and the ac generator playing the role of the noise source were attached to same sample contacts(see Fig.1, contacts 1,2). The experimental data demonstrate that the amplitude of the rectified voltage is $proportional$ to the ac voltage applied by the generator. This finding is rather puzzling as regards the rectification consept proposed in Refs.[3,5]. In the present paper, we suggest a phenomenological analysis of the experimental findings and then explain the important specific features of the effect.

\section{\label{sec:phenomenology1} Measurements at B=0}
\begin{figure}
\begin{center}
\includegraphics[scale=0.75]{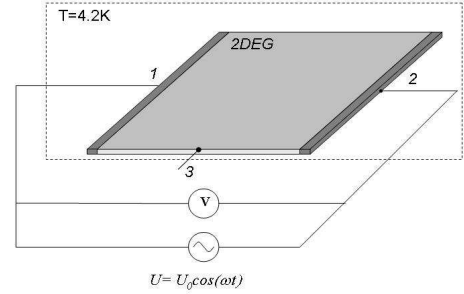} \caption[]{\label{Fig1} Setup configuration. The noise is simulated by the ac generator.}
\end{center}
\end{figure}

We further use the simplest model of a current-voltage characteristic of the 3D-2D Schottky contact[7,8]. In the thermionic diode approximation at finite temperatures, the current is given by
\begin{equation}
I= I_{0}\left(\exp\left( \frac{eV}{kT} \right )-1\right),
\label{Schottky diode}
\end{equation}
where $I_{0}=5.36A(kT)^{2}w\exp\left( \frac{\epsilon_{F}-eV_{0}}{kT} \right )$ is the backward saturation current; $V_{0}$, equilibrium contact potential; and, $V$, voltage drop across the contact. Then, $A=\frac{em*}{2\pi^{2}\hbar^{3}}$ is the Richardson constant for the thermionic emission, and $w$ is the quantum well width.

We emphasize that the Schottky diodes at the left( Fig.1, index 1 ) and right( index 2 ) contacts have the opposite polarity and, in general, are different from each other. Therefore, the relationship between the total voltage drop across the sample, $U$, and the current, $I$, reads:
\begin{equation}
u = ir+\ln \left( \frac{1+i}{1-ia} \right ),
\label{VAC}
\end{equation}
where $u=\frac{eU}{kT}$ is the dimensionless voltage; $i=I/I_{01}$, current scaled with respect to reverse saturation current($I_{01}$) of the left-contact diode; and, $a=I_{01}/I_{02}$, asymmetry parameter of the contacts. Then, $r=\frac{I_{01}Re}{kT}$ is the dimensionless resistance of the 2D electron gas, and $R$ is the 2DEG resistance.

\begin{figure}
\begin{center}
\includegraphics[scale=0.75]{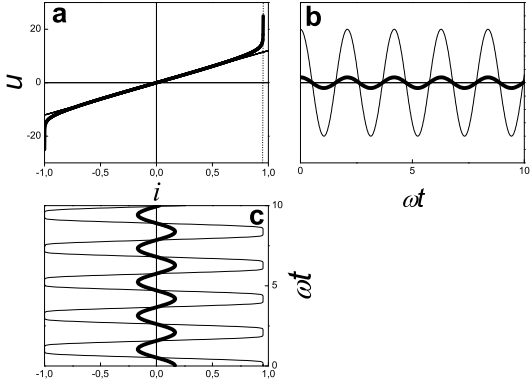} \caption[]{\label{Fig2} Current-voltage characteristics(panel a) specified by Eq.(2) for a contact-asymmetry parameter $a=1.05$ and a 2DEG resistance $r=10$. The dashed line represents the low-field ohmic dependence. The low-voltage(bold line) and high-voltage(thin line) ac input signals are represented in panel b. The respective responses are shown in panel c.}
\end{center}
\end{figure}

We are primarily interested in the low-current case $i \ll 1$ and, therefore, linearize Eq.(2) with respect to the current as
\begin{equation}
u = (1+a+r)i-\frac{1-a^{2}}{2}i^{2}+\frac{1+a^{3}}{3}i^{3}+...
\label{linear_u}
\end{equation}
As expected, the current-voltage characteristic exhibits the ohmic behavior $u=(1+a+r)i$ or $U=IR_{tot}$, where $R_{tot}=R_{1}+R_{2}+R$ is the zero-field total resistance of the sample, and $R_{1,2}=\frac{kT}{eI_{01,2}}$ are the Schottky resistances of contacts 1 and 2, respectively. In the opposite case of a high applied voltage, the current-voltage characteristic is strongly nonlinear. Indeed, in this case, the forward and reverse currents are limited( see Fig.2,a ) by the Schottky diode saturation currents $I_{02}$ and $I_{01}$, respectively.

Let us first seek the response of the 3D/2DEG/3D system to an applied ac voltage $u=u_{0}\cos(\omega t)$. The ac voltage is provided by the generator shown in Fig.1. At low voltages $u \ll 1$, Eq.(3) allows one to extract the current as
\begin{equation}
i = \frac{u}{1+a+r}+\frac{1-a^2}{2(1+a+r)^3}u^{2}+\beta u^{3}+...,
\label{linear_i}
\end{equation}
where $\beta=\frac{(1-a^{2})^{2}}{2(1+a+r)^5}-\frac{(1+a^{3})}{3(1+a+r)^4}$. According to Eq.(4), the in-phase response of the system to the applied ac voltage consists of the ohmic contribution and an, additional part, because $i=(u_{0}/(1+a+r)+3/4\beta u_{0}^{3})\cos(\omega t)$. Thus, we conclude that the widely used lock-in ac measurement method give [3] a sample resistance that is somewhat different from zero-field sample resistance $R_{tot}$.

We now intend to resolve the primary problem formulated in the present paper. Let us investigate the dc response of the circuit(see Fig.1) to an applied ac voltage. We emphasize that the second-order in voltage term in Eq.(4) describes the rectification properties of the 2DEG sample at $a \neq 1$. Equation 4 yields the time-averaged current $\overline {i}=\frac{\omega}{2\pi}\int_{0}^{\frac{2\pi}{\omega}}idt \ll 1$ and, then, the voltage drop measured by dc voltmeter is given by:
\begin{equation}
\overline {u}=\frac{(1-a^2)}{4(1+a+r)^2}u_{0}^{2},
\label{average-low}
\end{equation}
The polarity of the measured dc voltage is determined by the contact asymmetry. As expected, the transmission characteristic $\overline {u}(u_{0})$ is quadratic law at $u_{0} \ll 1$. In the opposite case of a strong ac excitation $u_{0} \gg 1$, the dc response can be found qualitatively with the help of Fig.2,c. Indeed, the rectified current can be regarded as a rectangular meander sequence with linear fronts. The higher the applied ac voltage, the sharper the front of the current pulse. After simple averaging, we obtain the dc current and, finally, the rectified voltage as
\begin{equation}
\overline{u}=\frac{(1-a)(1+a+r)}{2a}\left [1- \frac{(1+a+r)(1+a)}{\pi u_{0}a}\right ].
\label{average-high}
\end{equation}
At a high input ac signal $u_{0}\gg 1$, the measured voltage saturates $\overline {u}_{sat}=\frac{1}{2} (1/a-1)(1+a+r)$. One could expect that, at a moderate ac signal level $u_{0}=i(1+a+r), i \sim 1$, the low and high ac input cases merge, and, consequently, there could exist a certain part of the transmission characteristic which could be associated with a linear dependence[3,5].

To confirm our qualitative predictions, we present in Fig.3 the result of our numerical calculations. We use current-voltage characteristic specified by Eq.(2). Reversing this equation with respect to current, we find numerically the dependence $i(u)$. The successive averaging of the current caused by the input ac signal gives the related dc voltage drop across the sample, and, hence the transmission characteristic. As expected, the transmission characteristic follows the asymptote given by Eq.(5) at low excitations. At a high-level ac input $u_{0} \gg 1$, the dependence $\overline {u}(u_{0})$ can be approximated with Eq.(6), and then levels-off. For intermediate voltages $u_{0} \sim r=10 $, the transmission characteristic exhibits a nearly linear behavior $\overline{u}=-A+Bu_{0}$ in accordance with the experimental findings[5].

\begin{figure}
\begin{center}
\includegraphics[scale=0.75]{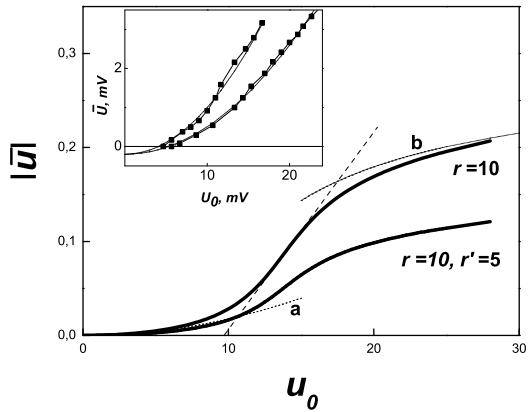} \caption[]{\label{Fig3} Main panel: the transmission characteristic for contact asymmetry $a=1.05$ and 2DEG resistance $r=10$(upper curve). Dotted lines a and b represent the low and high-voltage approximations specified by Eq.(5) and Eq.(6), respectively. The dashed line demonstrates a nearly linear dependence reported in [5]. The lower curve corresponds to the case of voltmeter connected to 1-3 contacts (see Fig.1) when $r=10, r'=5$. Inset: The observed [5] transmission characteristics correspond to those in the main panel.}

\end{center}
\end{figure}

It is instructive to note that, in Ref.[5], the dc voltmeter was connected to the 1st(basic) and  intermediate probe contacts (see contact 3 in Fig.1) as well. In this case, the transmission characteristic must be multiplied by the factor $\frac{1+a'+r'}{1+a+r}$, where $a'=I_{01}/I_{03}$ is the asymmetry of the intermediate contact with respect to the 1st one, and $r'$ is the partial resistance of the 2DEG related to the intermediate contact position. We emphasize that the lower the partial resistance $r'$, the smaller the differential slope $\frac{d\overline{u}}{d u_{0}}$ of the transmission characteristic(see the lower curve in Fig.3, main panel).

We now estimate the actual parameters of 3D/2DEG/3D system experimentally studied in Ref.[5]. For n-AlGaAs/GaAs sample( 2DEG density $n=3.46 \times 10^{11}$cm$^{-1}$, dielectric constant $\epsilon=12.7$, effective mass $m=0.068m_{e}$ ) we find the Fermi energy as $\epsilon_{F}=80$meV, whereas the Bohr energy $\epsilon_{B}=\frac{me^{4}}{2\kappa^{2}\hbar^{2}}=6.7$meV. As demonstrated in Ref.[9], the thermionic diode approximation is well justified when $T>T_{0}$, where $T_{0}=\frac{\epsilon_{F}}{k}\sqrt{\frac{\epsilon_{B}}{e(V_{0}-V)}}$. At $T<T_{0}$, the tunneling current across the Schottky diode becomes higher than the thermionic current. For the typical equilibrium contact potential $V_{0}=1$eV and zero diode bias $V=0$, we obtain $T_{0}=73$K, and, hence, the observed data[5] cannot be analyzed directly in terms of the thermionic mechanism[7]. Nevertheless, even at low temperatures the current-voltage characteristic behaves similar to that described by Eq.(1). Hence, the main assumption that the nonzero voltage results from rectification of the ac noise remains justified.

We now estimate the resistance of the Schottky contact. Using the data from [5], we find the total low-field resistance of the sample $R_{tot}=13k\Omega$. Then, in the insert of Fig.3 we reproduce the transmission characteristic data[5] for the dc output measured across 1-2(upper curve) and 1-3(upper curve) contacts. Both  curves demonstrate a puzzling threshold behavior, which can be, in principle, attributed to the possible voltmeter zero-point shift. Consequently, the low-voltage part of these curves can be approximated by the following equations $\overline{U}_{12}[V]=-0.0002+12U^{2}_{0}[V]$ and $\overline{U}_{13}=-0.0002+7U^{2}_{0}$ respectively. Note that, irrespective of the actual form of $\overline{u}(u_{0})$ dependence, these curves differ by the factor $\frac{R_{1}+R_{3}+R/2}{R_{1}+R_{2}+R}$ which is equal to the ratio $7/12$ found above by fitting. Assuming that the contact asymmetry is small, i.e., $R_{1}\sim R_{2}\sim R_{3}$  we obtain the Schottky contact resistance $R_{1}=1.05k\Omega$.

\section{\label{sec:phenomenology2}Zero oscillations in strong magnetic fields}
We argue that the zero-oscillations observed in 2DEG in strong magnetic fields originate from the noise rectification as well. Indeed, at a fixed magnetic field strength, the transmission characteristic observed [5] for the ZO amplitude is analogous to that reported for $B=0$. We remind that the sign of the rectified voltage depends on the asymmetry of the Schottky diode contact pair. If the the reverse current $I_{0}$ of the Schottky contact oscillates in the magnetic field, the rectified dc voltage(ZO) will oscillate as well. The detailed examination of the ZO oscillations will be made elsewhere.

\section{\label{sec:conclusion}Conclusion}

We demonstrate that the dc voltage observed at low-temperatures in a 2D electron sample $without$ external excitation is caused by the noise rectification by Schottky diodes formed at the sample contacts. At low noise level the rectified voltage as a function of the noise amplitude follows the usual quadratic law. At higher noise magnitudes, the rectified voltage exhibits a nearly linear behavior, and, finally saturates. The rectified voltage is shown to depend on the contact pair asymmetry. We suggest that the shunting of the sample contacts by a capacitance is a powerful tool for suppressing the rectified voltage.

The author wish to thank Prof. M.Dyakonov for helpful comments.

\end{document}